\documentclass[sigconf]{acmart}

\usepackage[utf8]{inputenc} 
\usepackage[T1]{fontenc}    
\usepackage{hyperref}       
\usepackage{url}            
\usepackage{booktabs}       
\usepackage{amsfonts}       
\usepackage{nicefrac}       
\usepackage{microtype}      
\usepackage{xcolor}         
\usepackage{graphicx}
\usepackage{bbding}
\usepackage{multirow}
\usepackage{graphicx}
\usepackage{subcaption}
\usepackage{tabularx}
\usepackage{tcolorbox}
\usepackage{lipsum}
\usepackage{natbib}

\usepackage{xcolor}
\usepackage{ifthen}

\newboolean{highlightmode}
\setboolean{highlightmode}{false}  

\usepackage{soul}  
\soulregister\ref7  
\soulregister\cite7  
\soulregister\pageref7  
\newcommand{\newadd}[1]{%
    \ifthenelse{\boolean{highlightmode}}{%
        \sethlcolor{yellow}\hl{#1}
    }{%
        #1
    }%
}

\usepackage{listings}
\lstdefinestyle{go}{
    language=Go,
    frame=lines,
    basicstyle=\ttfamily\small,
    keywordstyle=\color{blue},
    commentstyle=\color{teal},
    stringstyle=\color{red},
    numbers=left,
    numberstyle=\tiny\color{gray},
    stepnumber=1,
    numbersep=5pt,
    showstringspaces=false,
    breaklines=true,
    breakatwhitespace=true,
    tabsize=4,
    captionpos=b
}
\AtBeginDocument{%
  }

\setcopyright{none} 
\renewcommand\footnotetextcopyrightpermission[1]{}

\setcopyright{acmlicensed}
\copyrightyear{2018}
\acmYear{2018}
\acmDOI{XXXXXXX.XXXXXXX}

\acmConference[Conference acronym 'XX]{Make sure to enter the correct
  conference title from your rights confirmation email}{June 03--05,
  2018}{Woodstock, NY}
  
\acmISBN{978-1-4503-XXXX-X/2018/06}

\begin{document}

\title{LLM-based Agents for Automated Bug Fixing: How Far Are We?}


\author{Xiangxin Meng}
\affiliation{%
  \institution{ByteDance}
  \city{Beijing}
  \country{China}}
\email{mengxiangxin.1219@bytedance.com}

\author{Zexiong Ma}
\affiliation{%
  \institution{ByteDance}
  \city{Beijing}
  \country{China}}
\email{mazexiong@bytedance.com}

\author{Pengfei Gao}
\affiliation{%
  \institution{ByteDance}
  \city{Beijing}
  \country{China}}
\email{gaopengfei.se@bytedance.com}

\author{Chao Peng}
\authornote{Corresponding author.}
\affiliation{%
  \institution{ByteDance}
  \city{Beijing}
  \country{China}}
\email{pengchao.x@bytedance.com}

\renewcommand{\shortauthors}{Meng et al.}

\begin{abstract}

Large language models (LLMs) and LLM-based Agents have been applied to fix bugs automatically, demonstrating the capability in addressing software defects by engaging in development environment interaction, iterative validation and code modification.
However, systematic analysis of these agent systems remain limited, particularly regarding performance variations among top-performing ones.
In this paper, we examine six repair systems on the SWE-bench Verified benchmark for automated bug fixing.
We first assess each system's overall performance, noting the instances solvable by all or none of these systems, and explore the  capabilities of different systems.
We also compare fault localization accuracy at file and code symbol levels and evaluate bug reproduction capabilities.
Through analysis, we concluded that further optimization is needed in both the LLM capability itself and the design of Agentic flow to improve the effectiveness of the Agent in bug fixing.
\end{abstract}

\maketitle

\section{Introduction}
Large Language Models (LLMs)~\cite{zhao2023survey} are advanced machine learning models trained on vast amounts of textual data, capable of understanding and generating human-like text.
LLM-based Agents~\cite{xi2023rise} are systems that utilize large language models to interact with the environment and accomplish specific tasks.
Recently, LLM-based Agents have demonstrated significant influence in automated bug fixing in code repositories~\cite{kang2023explainable, zhang2024autocoderover, yang2024swe}.
Thanks to the powerful natural language processing capabilities of LLMs, 
these agents can efficiently understand and analyze source code and its associated natural language descriptions, 
such as user-submitted issue descriptions and code comments. 
Additionally, through dynamic interaction with local environments (e.g., via terminal), 
LLM-based agents can retrieve useful information from the code repository, perform code editing and execution, 
and iteratively validate repair results, 
thereby improving the accuracy and efficiency of bug fixes. 
This combination of LLM and environmental feedback has made bug fixing more efficient and feasible than ever before, 
providing revolutionary new tools for software maintenance and development. 

Researchers from both  industry~\cite{liu2024marscode, ma2024understandsoftwarerepository, blackbox, wandb, emergent, devlo, midwit} and academia~\cite{zhang2024autocoderover,tao2024magis, learn-by-interact} have developed LLM-based agent systems to locate and fix bugs in code repositories. 
To evaluate the fault localization and repair capabilities of LLMs and various agent systems, 
Jimenez et.al~\cite{jimenez2023swe} proposed the evaluation datasets SWE-bench, with derived versions SWE-bench Lite (a subset of the full benchmark), and SWE-bench Verified (human annotated subset of SWE-bench published recently). 
These datasets contain real bugs from code repositories and can verify the correctness of the patches generated by Agents through unit tests. 
Recently, these datasets have become the most influential benchmarks in the field of automated bug fixing, 
attracting both academic and industrial participants to compete on the corresponding leaderboard\footnote{\url{https://www.swebench.com/}}, with new submissions typically every one or a half week.

However, no work has systematically analyzed the fault localization and repair capabilities of LLM-based Agents or the performance differences among various tools within these Agent systems. 
Regarding the SWE-bench Verified dataset itself, 
due to the quality of \newadd{issue descriptions}\footnote{\newadd{An issue description is a concatenated string of "Issue Titles" and "Descriptions".}} and the complexity of the logical dependencies related to the defects, 
some instances in the benchmark are easier for Agents to fix, while others are more difficult~\cite{xia2024agentless}. 
As for the design of the systems, different designs exhibit different planning, reasoning, and problem solving capabilities, 
i.e. some systems adopting static approaches~\cite{gru} and others adopting dynamic approaches~\cite{zhang2024autocoderover}.
We have also observed significant differences in cases that each system can solve. 
Therefore, analyzing the solving capabilities of LLM-based Agents on specific instances can not only help us 
better understand the current performance of Agents but also provide comparative insights for future research directions.

We collected the six systems (i.e. W\&B Programmer~\cite{wandb}, Blackbox AI Agent~\cite{blackbox}, CodeStory Midwit Agent~\cite{midwit}, Learn-by-interact~\cite{learn-by-interact}, Devlo~\cite{devlo}, Emergent E1~\cite{emergent} with top performances on the SWE-bench Verified Leaderboard and 
conducted a comprehensive analysis of the performance differences of each system.  
First, we evaluated the overall performance of LLM-based Agents in bug fixing, 
including statistics on the instances that all six systems can solve and those that none can solve, 
and analyzed the reasons behind these results. 
Next, we investigated the performance differences in fault localization among different systems and their causes, 
compiling file-level and code symbol-level localization accuracy rates. 
Finally, we analyzed the impact of bug reproduction on bug fixing.

Through data analysis, we have summarized several insights. To improve bug fixing, it is essential to enhance the model's reasoning ability, enabling it to accurately identify bug-related information within an issue and reduce noise interference. For multiple potential repair locations, the model should leverage its reasoning capabilities to determine the most relevant location. From the Agentic flow's perspective, there should be a strong focus on the quality of the issue and attention to multiple suspicious locations in the stack trace. The design should include mechanisms to verify the completeness of patches and consider their global impact. Mechanisms should also be implemented to mitigate the randomness of the model's output or effectively utilize its diversity. In fault localization, achieving code symbol-level accuracy is more critical than file-level and line-level. During the reproduction process, ensuring the correctness of reproductions is crucial, as incorrect reproductions can result in the failure of the entire solving process.

\noindent\textbf{Contributions:} To the best of knowledge, this is the first work to:
\begin{itemize}
    \item Study the effectiveness of LLM-based Agents in automatic bug fixing for code repositories.
    \item Examine the effectiveness of different LLM-based Agents in Fault Localization and analyze their differences.
    \item Investigate the impact of bug reproduction on bug fixing of LLM-based Agents.
    \item Summarize the current issues and future research directions for LLM-based Agents in bug fixing.
\end{itemize} 

\noindent\textbf{Paper Organization }
The remainder of this paper is organized as follows: 
Section~\ref{sec:background} explains the background. 
Section~\ref{sec:study_design} describes the study design.
Section~\ref{sec:analysis_results} presents the analysis results and findings. 
Section~\ref{sec:discussion} discusses the analysis results and findings. 
Section~\ref{sec:threats} reports the threats to validity. 
Section~\ref{sec:related} discusses related work, and Section~\ref{sec:conclusion} concludes the paper.

\section{Background}
\label{sec:background}
In this section, we first introduce SWE-bench benchmark and then we introduce the leading LLM-based bug fixing systems.
\subsection{SWE-bench Benchmark}
SWE-Bench~\cite{jimenez2023swe} is a comprehensive benchmark designed to evaluate LLMs on complex real-world software engineering tasks sourced from GitHub issues and corresponding pull requests across 12 popular Python repositories.
This benchmark addresses the limitations of existing coding benchmarks such as HumanEval~\cite{chen2021evaluating} by presenting tasks that require models to understand and coordinate changes across large codebases involving multiple functions and files.
The benchmark includes 2,294 task instances and emphasizes the need for models to interact with execution environments and handle long contexts, showcasing the challenges that real-world software engineering problems pose to current LLMs.
Their evaluations reveal that even the best-performing models at the time of publication, such as Claude 2, achieve a success rate of only 1.96\%, highlighting significant room for improvement.

After the release of the benchmark, OpenAI's testing identified several challenging or possibly unsolvable tasks within SWE-bench benchmark, which may cause the benchmark to systematically underestimate the autonomous software engineering capabilities of models. To address these issues, OpenAI collaborated with the authors of SWE-bench to develop a new release of the benchmark that should provide more accurate evaluations, named SWE-bench Verified\footnote{\url{https://openai.com/index/introducing-swe-bench-verified/}}, which contains 500 instances with higher quality. It retains the diversity of SWE-bench but is easier to evaluate. After the release of SWE-bench Verified, it quickly became the most popular benchmark for evaluating large language models and Agent systems on coding performance.


\subsection{Leading LLM-based Bug Fixing Systems}

LLM-based Bug Fixing Systems are systems built on Large Language Models (LLMs) that can automatically edit code repositories to fix bugs based on issue reports. Bug fixing is a highly resource-intensive task in software development, requiring developers to reproduce the bugs reported in issue reports, precisely locate defective code snippets within large code repositories, understand the cause of errors, and implement fixes. Automating bug fixing has long attracted widespread attention in both academia and industry. Given the strong logical reasoning and coding capabilities demonstrated by LLMs, numerous works have explored the development of automated bug fixing tools based on LLMs. In this paper, we study six leading LLM-based Bug Fixing Systems, comparing their differences in system design and performance in automated bug fixing, analyzing the shortcomings and limitations of existing systems, and providing direction for future work in building adaptive, high-reliability automated bug fixing systems. 
Below this section, we will introduce these systems from a technical perspective:

\textbf{W\&B Programmer~\cite{wandb}}, developed by wandb, leverages OpenAI's o1 model with a high reasoning mode for each step and editing decision. It incorporates a GPT-4o based memory component to compress the agent's history, a custom Python code editor optimized for model context, and the ability to register "auto-commands" that execute after each edit. To enhance performance, the agent employs five parallel rollouts per instance, followed by a "crosscheck" step using o1 as a tie-breaker to select the best rollout.

\textbf{Blackbox AI Agent~\cite{blackbox}} developed by Blackbox, is a sophisticated tool designed to enhance productivity and streamline workflows. The agent is equipped with a suite of powerful tools, including \texttt{get\_repo\_structure}, \texttt{open\_file}, \texttt{create\_file}, \texttt{search\_file}, \\
and the ability to execute \texttt{bash commands}. These tools enable the agent to efficiently navigate, manipulate, and manage code repositories and files. LLMs leverage these tools in a ReAct (Reasoning and Acting) format to perform a variety of tasks. It can filter and extract relevant content, generate file paths, and validate patches by executing tests. This approach ensures that the agent not only retrieves the necessary information but also verifies the integrity and functionality of the code through rigorous testing.

\textbf{CodeStory Midwit Agent~\cite{midwit}}, an agent released by CodeStory. They incorporate a suite of specialized tools, including \texttt{list\_files}, \texttt{open\_files}, \texttt{str\_replace\_editor}, \texttt{attempt\_completion}, \\
\texttt{ripgrep\_search}, and \texttt{terminal\_access} for Agent. It utilizes Mon-te Carlo Tree Search\cite{antoniades2024swe} to ensure efficient exploration of the candidate space and filters trajectories by calculating the mean reward of the trajectory's steps. This approach enhances the performance of the Agent system on SWE-Bench tasks.

\textbf{Learn-by-interact~\cite{learn-by-interact}} employs an agent that directly interacts with code repositories in offline mode to gather extensive high-quality interaction trajectories. It leverages in-context learning to strategically integrate these historical interaction patterns into prompts. This enables the agent to systematically reference and adapt prior successful debugging procedures when addressing novel software defects, resulting in progressively improved bug resolution effectiveness through accumulated experiential knowledge.

\textbf{Devlo~\cite{devlo}}, an agent released by Devlo. Devlo designs tools such as \texttt{exploring\_repository\_structure}, \texttt{creating\_file}, \texttt{modifying code}, and \texttt{run\_script}. Additionally, Devlo employs rigorous testing mechanisms such as \texttt{reproduction\_test} and \texttt{edge\_case\_test} to meticulously filter and validate candidate patches, ensuring robust and reliable outcomes.

\textbf{Emergent E1~\cite{emergent}} initially approaches the problem by engaging multiple expert agents, each equipped with tools such as \texttt{viewing}, \texttt{creating\_file}, and \texttt{str\_replace}. Subsequently, the final patch is selected from among multiple candidate patches.

In this paper, we will analyze the localization accuracy of different localization strategies and the impact of different components on the final patch generation accuracy, providing guidance for building more reliable bug fixing systems in future work.

\section{Study Design}
\label{sec:study_design}
In this section, we introduce the research questions and the design of issue quality assessment criteria.
\subsection{Research Questions}
\noindent\textbf{RQ1. Effectiveness of Systems: How does the LLM-based Agent currently perform in automatic bug fixing in code repositories?}

\noindent\textit{\textbf{\underline{Motivation:}}} In the SWE-bench Verified leaderboard, resolve rates of different systems vary significantly, with substantial differences in instances each system can and cannot solve, often due to issue description quality and system design. Analyzing why high-quality descriptions sometimes fail while low-quality ones succeed is necessary.

\noindent\textit{\textbf{\underline{Approach:}}} We will analyze differences in instances solved by various systems, showing how many are resolved by all systems and how many by none. Similar to the assessment done by Agentless on the SWE-bench Lite dataset, we will design an automated method for quality assessment based on state-of-the-art large language models for the SWE-bench Verified dataset. We will investigate why many high-scoring issues are unresolved by any system, while some low-scoring issues are resolved by all systems.

\noindent\textbf{RQ2. Effectiveness of FL: How do different systems perform in Fault Localization and what are the reasons for their differences?}

\noindent\textit{\textbf{\underline{Motivation:}}} Fault localization is crucial in bug fixing since accurate localization increases the probability of fixing the bug. We need to investigate the differences in fault localization effectiveness among different systems.

\noindent\textit{\textbf{\underline{Approach:}}} Based on the ground truth, we will compile statistics on the proportion of successfully localized faulty files and code symbols (including classes, functions, methods, and top-level statements) for each system in each SWE-bench Verified instance.

\noindent\textbf{RQ3. Effectiveness of Reproduction: How effective is the bug reproduction task currently?}

\noindent\textit{\textbf{\underline{Motivation:}}} Bug reproduction is vital in bug fixing and dynamic debugging. It aids fault localization through error messages and validates patches. Higher-quality reproduction scripts provide more accurate information, increasing the probability of fixing the bug. Investigating the effectiveness of bug reproduction is crucial.

\noindent\textit{\textbf{\underline{Approach:}}} Due to the difficulty of collecting bug reproduction scripts from various repair systems, we designed an Agent repair system with a mainstream repair process. We collected the content and execution commands of reproduction scripts, subsequently evaluating their quality based on output differences before and after applying the golden patch.





\subsection{Issue Quality Assessment Criteria}
We design a scoring mechanism based on five metrics, allowing us to evaluate the quality of different issue descriptions across multiple dimensions. Specifically, we drew inspiration from the five metrics used in Agentless~\cite{xia2024agentless} for evaluating the quality of issue descriptions: Quality of Bug Locations (File/Function/Line), Quality of Reproducible Examples, and Quality of Resolve Solutions. Agentless systematically evaluated the quality of issue descriptions for 300 instances in SWE-bench Lite benchmark using these five metrics. Since SWE-bench Verified was recently proposed, there has been no research conducting similar evaluations on it so far. To fill this gap, we designed clear scoring criteria and utilized the DeepSeek-R1\footnote{\url{https://www.deepseek.com/}} model, one of the models with the best reasoning abilities and cost-effectiveness, to implement the five-dimensional quality assessment for instances in SWE-bench Verified. One point to note is that the Function-level Fault Localization granularity proposed in Agentless was replaced with a similar but broader granularity of code symbol-level Fault Localization in our study, where code symbols include classes, functions, class methods, and top-level statements.
\newadd{We explain the condensed scoring criteria for the five dimensions as follows, and full prompts for the LLM to make the quality assessments can be obtained in our homepage\footnote{\url{https://github.com/ResearchOpenRepos/bug_fixing_agent_empirical_study/tree/master/issue_quality_annotation}}.}

\textbf{Quality of File-level Fault Localization:} We identify edited code files based on the golden patch (human-written patch, considered ground truth) and treat them as \textit{golden files}. We then input the \textit{issue description} and the corresponding \textit{golden files} to the LLM to determine the extent to which golden files appear in issue descriptions. The scoring criteria are:

\begin{enumerate}
    \item \textbf{StackTrace (3)}. \newadd{The issue description contains stack trace information, which contains the modified file name.}
    \item \textbf{Keyword (2)}. \newadd{The issue description contains the file name and its path prefix. The path prefix is required to uniquely identify the file entity. For example, for \textit{path/to/file\_reader.py}, either \textit{path/to/file\_reader.py} or \textit{path/to/file\_reader} is acceptable. File path and name may appear in different locations but both must be present in the issue description.}
    \item \textbf{Natural Language (1)}. \newadd{The issue description provides the file name in natural language without path prefix. The file name cannot be split - \textit{file\_reader.py} or \textit{file\_reader} is valid, but \textit{file} and \textit{reader} appearing separately does not qualify.}
    \item \textbf{No Information (0)}. \newadd{This applies when keywords appear separately without forming the complete file name, for example, \textit{file} and \textit{reader} appear individually rather than as \textit{file\_reader}, or when no relevant keywords appear.}
\end{enumerate}

\newadd{Since a single case may contain multiple golden files, the LLM is required to evaluate each file based on the aforementioned scoring criteria, with the maximum score ultimately being selected as the annotation result for that case.}

\textbf{Quality of Symbol-level Fault Localization:} \newadd{Similar to file-level localization, we extract modified code symbols (classes, functions, methods) from golden patches via AST analysis, storing them in \textit{file\_path::symbol\_name} format and providing them to the LLM with issue descriptions. As a case may also involve multiple symbol modifications, its final score is the maximum of all symbol-level evaluations. The scoring criteria are:}

\begin{enumerate}
    \item \textbf{StackTrace (3)}. \newadd{The issue description contains stack trace information, which contains the target code symbol name.}
    \item \textbf{Keyword (2)}. \newadd{The issue description includes both the symbol name and its associated file path. While a code symbol may have multiple definitions across different files, the \textit{file\_path::symbol\_name} combination enables more accurate localization. Notably, we relax the requirement for \textit{complete} file paths since the presence of code symbols in issue descriptions itself provides sufficient contextual constraints.}
    \item \textbf{Natural Language (1)}. \newadd{The issue description mentions the code symbol name in natural language without file path. Following the same principle as file-level localization, these symbol names must remain intact as atomic units, e.g., the symbol \textit{get\_name} would not be considered matched if only its constituent parts (\textit{get} or \textit{name}) appear independently.}
    \item \textbf{No Information (0)}. \newadd{The issue description contains neither direct nor sufficiently indicative references to the target symbol. Partial or indirect mentions (e.g., separate components like \textit{get} and \textit{name} appearing without the complete symbol \textit{get\_name}) are treated as no symbol information.}
\end{enumerate}


\textbf{Quality of Line-level Fault Localization:} We use an out-of-the-box tool named delta\footnote{\url{https://github.com/dandavison/delta}} to convert the standard diff format golden patch into a \textit{show-line golden patch}, displaying line numbers before and after the patch at the beginning of each line. This annotation aids the model in understanding code modifications and accurately matching line numbers. We then provide the issue description and the show-line golden patch to the LLM. The scoring criteria are:

\begin{enumerate}
    \item \textbf{StackTrace (3)}: \newadd{The description must contain a stack trace that explicitly includes at least one pair of <file\_name, line\_number> corresponding to the modified code locations in the patch.}
    \item \textbf{Keyword (2)}: \newadd{At least one pair of <file\_name, line\_number> appears in the issue text, but not appears in a stack trace.}
    \item \textbf{Natural Language (1)}: \newadd{The issue description contains a natural language explanation of the buggy line(s), but does not specify the exact line number(s). For example, if the patch contains \textit{result = get\_content(url, proxy, timeout)}, the description including \textit{result = get\_content(url, proxy, timeout)} or \textit{get\_content(url, proxy, timeout)} qualifies. Isolated identifiers like \textit{proxy} or \textit{timeout} do not qualify.}
    \item \textbf{No Information (0)}: \newadd{No useful line-related information appears in the issue description.}
\end{enumerate}

\textbf{Quality of Reproducible Examples:} \newadd{We provide only the issue description, without additional ground truth information. Consequently, our approach only requires the LLM to detect the presence of reproducible examples, without evaluating their semantic correctness. The scoring criteria are:}

\begin{enumerate}
    \item \textbf{Contains REs (3)}. The issue description includes complete, directly runnable code to reproduce the problem without additional setup.
    \item \textbf{Contains Partial REs (2)}. The description provides partial code, containing the main part needed but lacking some information, requiring extra context or setup to run.
    \item \textbf{Info in NL (1)}. The description provides a detailed, step-by-step explanation in natural language on how to reproduce the problem, without code snippets.
    \item \textbf{Not Enough Info (0)}. There is insufficient information to reproduce the problem.
\end{enumerate}

\textbf{Quality of Resolve Solutions:} The model receives the issue description and the golden patch. Scoring criteria are:

\begin{enumerate}
    \item \textbf{Exact Patch (3)}. \newadd{The issue description provides a code solution that is exactly the same as the correct patch. If there is any semantic discrepancy or if the code snippet only covers part of the standard patch rather than the entire content, this option should not be selected.}
    \item \textbf{Complete Steps in NL (2)}. \newadd{The issue description provides a complete natural language solution that is semantically consistent with the correct patch. One could reasonably write the correct patch based on this description without additional guidance. If the description only covers part of the standard patch, this option should not be selected.}
    \item \textbf{Some Steps in NL (1)}. \newadd{The issue description provides a partial natural language solution that is not semantically complete compared to the provided patch. One could write part of the correct patch based on this description but not all. This occurs when users have incomplete understanding and provide partial solutions.}
    \item \textbf{No Solution (0)}. \newadd{The issue description does not provide any solution description - the user only points out the problem without suggesting how to fix it.}
    \item \textbf{Misleading Information (-1)}. \newadd{The issue description includes the user's guess on how to solve the problem, but this guess is completely wrong and off track compared to the correct patch, making it misleading information.}
\end{enumerate}

\subsection{Reliability of the Scoring System}

\newadd{
To validate the reliability of the scoring mechanism using DeepSeek R1, we performed a cross-validation against deterministic regex rules designed to identify structured cues such as stack traces in issue descriptions. Specifically, we analyzed all 500 issues and found disagreements in 34 cases. A detailed inspection revealed that DeepSeek substantially outperforms regex-based methods. Regex rules, though precise, often fail to generalize across diverse stack trace formats, which vary by Python version, bug-tracking tool, and runtime environment (e.g., Jupyter Notebook). Furthermore, many issue reports describe stack traces or error locations using natural language rather than standardized formatting. In such cases, DeepSeek’s semantic understanding allows it to successfully identify valuable debugging signals that regex rules miss. This result supports the use of DeepSeek R1 as a more robust and adaptable scoring mechanism for assessing issue quality.
}





\section{Analysis \& Results}
\label{sec:analysis_results}

We will sequentially present the analysis results and insights for RQ1 through RQ3.

\subsection{RQ1: Effectiveness of Systems}
\begin{figure*}
  \centering
  \includegraphics[width=\textwidth]{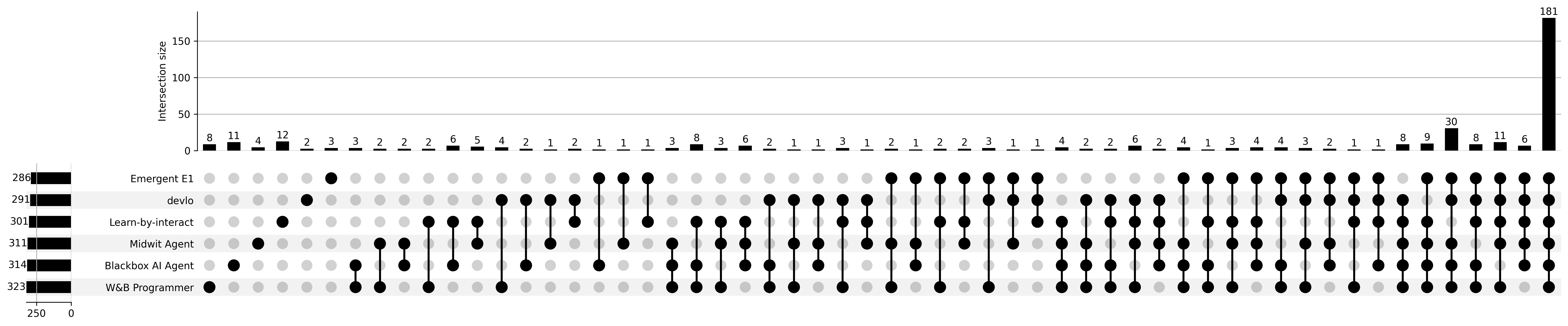}
  \caption{Analysis of Performance of State-of-the-art Techniques on Issue Resolve Task}
  \label{fig:repair_analysis}
\end{figure*}

We analyzed the versions of cases that each of the six tools can individually solve, as well as the differences between the cases that each tool can resolve, as shown in Figure~\ref{fig:repair_analysis}. The histogram at the top of the figure shows the number of cases each tool can resolve in the SWE-bench Verified dataset. Specifically, ranked from lowest to highest, these are Emergent E1, Devlo, Learn-by-interact, Midwit Agent, Blackbox AI Agent, W\&B Programmer, resolving 286, 291, 301, 311, 314, and 323 cases, respectively. W\&B Programmer performs the best, achieving a 13.0\% performance improvement over Emergent E1 and addressing 64.6\% of the total 500 cases in the SWE-bench Verified dataset. Compared to the popular APR benchmark Defects4J~\cite{just2014defects4j} over the past decade, SWE-bench Verified introduces stricter usage protocols, prohibiting participants from leveraging dynamic evaluation results generated by closely related failing test cases as feedback information for filtering patches. This test case set can only be utilized as a quality standard once the patch generation process has been completed. In this context, considerable of FL methods based on dynamic test execution information (e.g., spectrum-based and mutation-based methods) cannot be used, adding further difficulty to problem detection and resolution. This strict protocol undoubtedly aligns more closely with real-world development scenarios, where repair tools must rely almost solely on issues raised by users and the current state of the code repository to devise solutions. Against this backdrop, W\&B Programmer's ability to address 64.6\% of cases underscores its advanced capabilities and utility in real-world development environments.

    

In Figure~\ref{fig:repair_analysis}, \newadd{the histogram to the right of the tool names presents the count of case versions that different tool combinations can address. Each column indicates the number of case versions solvable by the tools marked with black dots but not by those marked with gray dots. For instance, in the first column, only W\&B Programmer is marked with a black dot, while the other five tools are marked with gray dots, indicating that W\&B Programmer can solve a unique set of 8 cases that none of the other five tools can handle. Similarly, the seventh column demonstrates that W\&B Programmer and Blackbox AI Agent together can solve 3 cases that the other four tools cannot address. The final column shows that there are 181 cases that can be solved by each of 6 tools.} 

\textbf{Analysis of Case Solvability.} Among the 500 cases in SWE-bench Verified, 404 cases are solvable by at least one of the six tools, while 96 cases remain unsolved by any tool. Furthermore, 181 cases can be solved by all six tools (represented by the last column in Figure~\ref{fig:repair_analysis}). We hypothesize that the issue descriptions for these 404 solvable cases are generally of higher quality, whereas the 96 cases that none of the tools can resolve likely exhibit lower-quality issue descriptions. To validate this hypothesis, we conducted a significance analysis of issue quality differences using five metrics. For quantitative comparison, we assigned a value to each option for each metric, as described in Section~\ref{sec:study_design} (e.g., 3 for StackTrace). The only point to note is that we assigned a score of -1 to the Misleading option in the Quality of Resolve Solutions metric. This ensures that a score of 0 consistently represents "no information" across all metrics. Hence, the scoring range was set from -1 to 3 instead of 0 to 4, making scores across different indicators comparable.

To examine the validity of our hypothesis, we calculated the mean values for the five indicators across two case sets: the set of 96 cases unsolved by any tool (referred to as the "No-one-resolve Set") and the set of 181 cases solvable by all tools (referred to as the "All-resolve Set"). These results are shown in Table~\ref{tab:issue_analysis}. For each indicator, a higher score indicates that the corresponding issue description provides more complete and detailed information. It is evident that the scores for all five indicators are consistently higher in the All-resolve Set compared to the No-one-resolve Set. This suggests that issue quality significantly influences the effectiveness of resolution methods, underscoring the importance of crafting clear and comprehensive issue descriptions to improve resolution rates from the outset.

\begin{table*}[htbp]
\centering
\caption{Issue Quality Analysis for All-resolve Cases and No-one-resolve Cases.}
\label{tab:issue_analysis}
\small
\begin{tabular}{c|c|c|c|c|c|c}
\hline
\textbf{Case Sets} & \textbf{\begin{tabular}[c]{@{}c@{}}Quality of\\ Reproducible\\ Examples\end{tabular}} & \textbf{\begin{tabular}[c]{@{}c@{}}Quality of\\ Resolve\\ Solutions\end{tabular}} & \textbf{\begin{tabular}[c]{@{}c@{}}Quality of\\ File-level\\ Locations\end{tabular}} & \textbf{\begin{tabular}[c]{@{}c@{}}Quality of\\ Symbol-level\\ Locations\end{tabular}} & \textbf{\begin{tabular}[c]{@{}c@{}}Quality of\\ Line-level\\ Locations\end{tabular}} & \multicolumn{1}{c}{\textbf{\begin{tabular}[c]{@{}c@{}}Avg\\ Scores\end{tabular}}} \\ \hline
\textbf{\begin{tabular}[c]{@{}c@{}}No-one-resolve\\ Set (96 cases)\end{tabular}} & 2.01 & 0.656 & 1.052 & 0.906 & 0.813 & 1.087 \\ \hline
\textbf{\begin{tabular}[c]{@{}c@{}}All-resolve Set\\ (181 cases)\end{tabular}} & 2.022 & 1.276 & 1.326 & 1.177 & 0.994 & 1.359 \\ \hline
\textbf{\begin{tabular}[c]{@{}c@{}}All-resolve Set / \\ No-one-resolve Set\end{tabular}} & 101\% & 195\% & 126\% & 130\% & 122\% & 125\% \\ \hline
\end{tabular}
\end{table*}

Examining the performance details for each indicator across both case sets, the most notable difference is observed in the \textit{Quality of Resolve Solutions}
metric, where the average score in the All-resolve Set is 1.95 times that of the No-one-resolve Set. This indicates that, at least within the SWE-bench Verified dataset, positive modifications based on observed symptoms can substantially aid in improving the possibility of solving the problem. This is followed by symbol-level, file-level and line-level bug locations, with score differences of 1.30, 1.26 and 1.22 times, respectively. 
The differences between these three metrics are minimal. Relatively, providing code symbol-level localization results might be more helpful.
Lastly, \textit{Quality of Reproducible Examples} shows only a 1.01-time score difference between the two sets, suggesting that the completeness of this indicator has a comparatively minor impact on enhancing the solvability of bug-fixing systems. 

In the above analysis, we validated our hypothesis by calculating differences across five metrics that represent issue quality in the All-resolve Set and No-one-resolve Set, showing a clear disparity in issue quality between cases that can be resolved and those that cannot. This reflects an overall trend, though some outliers may exist. On one hand, we observed several high-scoring cases in the No-one-resolve Set; we aim to investigate why, despite relatively complete information in these issues, no tools were able to resolve them. On the other hand, we also identified some low-scoring cases within the All-resolve Set, which may further highlight the capability of repair tools to independently find and gather necessary information. To this end, we collected the top 5 cases with the highest average scores across the five metrics in the No-one-resolve Set, and the 5 cases with the lowest average scores in the All-resolve Set, as shown in Table~\ref{tab:case_analysis}. It is evident that cases with a relatively high average score but cannot be successfully resolved generally have lower resolve solution quality. Hence, this feature appears to be the most critical factor influencing whether an issue can be resolved, which is consistent with our previous conclusions. For cases with relatively low average scores but can be resolved by all repair systems, their feature distributions vary significantly. Some provide different granular levels of bug location information, while others include reproducible examples and resolve solutions. A case-by-case analysis is needed to draw conclusions.

\begin{table*}[htbp]
\centering
\caption{Scores of 5 top-ranked cases in No-one-resolve Set and 5 bottom-ranked cases in All-resolve Set.}
\label{tab:case_analysis}
\small
\begin{tabular}{ccccccc}
\hline
\multicolumn{7}{c}{\textbf{No-one-resolve Set}} \\ \hline
\multicolumn{1}{c|}{\textbf{Case IDs}} & \multicolumn{1}{c|}{\textbf{\begin{tabular}[c]{@{}c@{}}Scores of\\ Reproducible\\ Examples\end{tabular}}} & \multicolumn{1}{c|}{\textbf{\begin{tabular}[c]{@{}c@{}}Scores of\\ Resolve\\ Solutions\end{tabular}}} & \multicolumn{1}{c|}{\textbf{\begin{tabular}[c]{@{}c@{}}Score   s of\\ File-level\\ Locations\end{tabular}}} & \multicolumn{1}{c|}{\textbf{\begin{tabular}[c]{@{}c@{}}Scores of\\ Symbol-level\\ Locations\end{tabular}}} & \multicolumn{1}{c|}{\textbf{\begin{tabular}[c]{@{}c@{}}Scores of\\ Line-level\\ Locations\end{tabular}}} & \textbf{\begin{tabular}[c]{@{}c@{}}Avg\\ Scores\end{tabular}} \\ \hline
\multicolumn{1}{c|}{\textbf{astropy-13977}} & \multicolumn{1}{c|}{3} & \multicolumn{1}{c|}{2} & \multicolumn{1}{c|}{3} & \multicolumn{1}{c|}{3} & \multicolumn{1}{c|}{3} & 2.8 \\ \hline
\multicolumn{1}{c|}{\textbf{astropy-14182}} & \multicolumn{1}{c|}{3} & \multicolumn{1}{c|}{0} & \multicolumn{1}{c|}{3} & \multicolumn{1}{c|}{3} & \multicolumn{1}{c|}{3} & 2.4 \\ \hline
\multicolumn{1}{c|}{\textbf{django-13513}} & \multicolumn{1}{c|}{2} & \multicolumn{1}{c|}{3} & \multicolumn{1}{c|}{2} & \multicolumn{1}{c|}{3} & \multicolumn{1}{c|}{2} & 2.4 \\ \hline
\multicolumn{1}{c|}{\textbf{django-16667}} & \multicolumn{1}{c|}{3} & \multicolumn{1}{c|}{0} & \multicolumn{1}{c|}{3} & \multicolumn{1}{c|}{3} & \multicolumn{1}{c|}{3} & 2.4 \\ \hline
\multicolumn{1}{c|}{\textbf{django-16938}} & \multicolumn{1}{c|}{3} & \multicolumn{1}{c|}{0} & \multicolumn{1}{c|}{3} & \multicolumn{1}{c|}{3} & \multicolumn{1}{c|}{3} & 2.4 \\ \hline
\multicolumn{7}{c}{\textbf{All-resolve Set}} \\ \hline
\multicolumn{1}{c|}{\textbf{django-10914}} & \multicolumn{1}{c|}{1} & \multicolumn{1}{c|}{0} & \multicolumn{1}{c|}{0} & \multicolumn{1}{c|}{0} & \multicolumn{1}{c|}{1} & 0.4 \\ \hline
\multicolumn{1}{c|}{\textbf{sympy-22456}} & \multicolumn{1}{c|}{0} & \multicolumn{1}{c|}{0} & \multicolumn{1}{c|}{1} & \multicolumn{1}{c|}{1} & \multicolumn{1}{c|}{0} & 0.4 \\ \hline
\multicolumn{1}{c|}{\textbf{django-13516}} & \multicolumn{1}{c|}{1} & \multicolumn{1}{c|}{0} & \multicolumn{1}{c|}{0} & \multicolumn{1}{c|}{0} & \multicolumn{1}{c|}{1} & 0.4 \\ \hline
\multicolumn{1}{c|}{\textbf{sphinx-8721}} & \multicolumn{1}{c|}{1} & \multicolumn{1}{c|}{0} & \multicolumn{1}{c|}{1} & \multicolumn{1}{c|}{0} & \multicolumn{1}{c|}{0} & 0.4 \\ \hline
\multicolumn{1}{c|}{\textbf{django-15499}} & \multicolumn{1}{c|}{0} & \multicolumn{1}{c|}{2} & \multicolumn{1}{c|}{0} & \multicolumn{1}{c|}{0} & \multicolumn{1}{c|}{0} & 0.4 \\ \hline
\end{tabular}
\end{table*}

We first conducted a manual inspection of the trajectories of the five high-scoring cases that could not be repaired and summarized the following insights:

\begin{tcolorbox}
 \textbf{Insights:} (1) Enhancing the understanding of root causes is crucial. The model’s understanding should extend beyond the location where the issue occurs (i.e., the symptoms) and include deeper reasoning about the relationship between multiple suspicious locations in the issue title or stack trace and the root cause. This would help prioritize the root cause location more effectively. (2) Improving the ability to generate patches for related locations and to verify patch completeness offers a feasible approach to further enhance issue resolution effectiveness from a holistic perspective.
\end{tcolorbox}

Subsequently, we conducted a manual inspection of the trajectories of 5 low-scoring cases that could be repaired by all 6 repair systems, and summarized the following insights:


\begin{tcolorbox}
 \textbf{Insights:} (3) The manually crafted patch may not be the only solution to the issue; multiple resolution strategies may exist. The model may attempt to generate a semantically correct or partially correct patch based on the suspicious locations already highlighted in the issue. (4) LLMs and agents demonstrate strong capabilities for discovering relevant information. They can deeply analyze the natural language issue description, iteratively extracting high-relevance code segments over multiple interactions.
\end{tcolorbox}

As shown in Figure~\ref{fig:repair_analysis},  we also discovered that Learn-by-interact~\cite{learn-by-interact} was able to independently repair 12 cases, the highest among the six repair systems. However, there were 30 cases that could be repaired by any of the other five systems, but not by Learn-by-interact, making it the most noteworthy in this aspect. 

Learn-by-interact uses an agent that interfaces with offline code repositories to collect extensive, high-quality interaction trajectories. It strategically integrates these historical interactions into prompts via in-context learning, enabling the agent to reference and adapt successful debugging methods to new software issues. This systematic approach results in progressively better bug resolution as experiential knowledge accumulates. In contrast, the other five systems focus on agent process design without leveraging prior experience, which is a relatively novel idea in Learn-by-interact. The significant design differences between the two sets of systems appear to lead to varied impacts on different cases, as the experiment results suggest. 

One insight is that generating patch candidates using diverse workflow designs within the process and the voting on the patches via their validity after generation, potentially enhancing the final effectiveness as the number of overall fixable cases increases.

\begin{tcolorbox}
 \textbf{Insights:} (5) Using designs that deviate from mainstream processes might help to uncover more repairable cases. For instance, leveraging experience gained from historical repair processes could aid in solving difficult instances. (6) Integrating different repair strategies into the entire workflow to generate a wider variety of patch candidates, and then voting on the validity of these patches, might enhance the final repair effectiveness.
\end{tcolorbox}

\begin{figure*}
  \centering
  \begin{subfigure}[b]{\textwidth}
    \includegraphics[width=\textwidth]{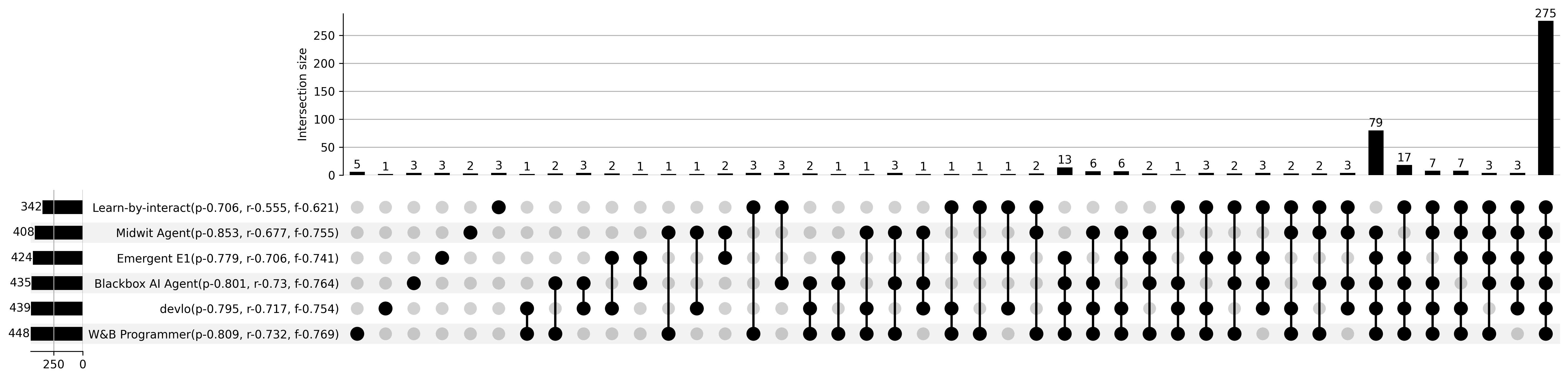}
    \caption{Statistics of samples where at least one buggy file is hit}
    \label{fig:file_fl_intersection}
  \end{subfigure}
  
  \vspace{1em}  
  
  \begin{subfigure}[b]{\textwidth}
    \includegraphics[width=\textwidth]{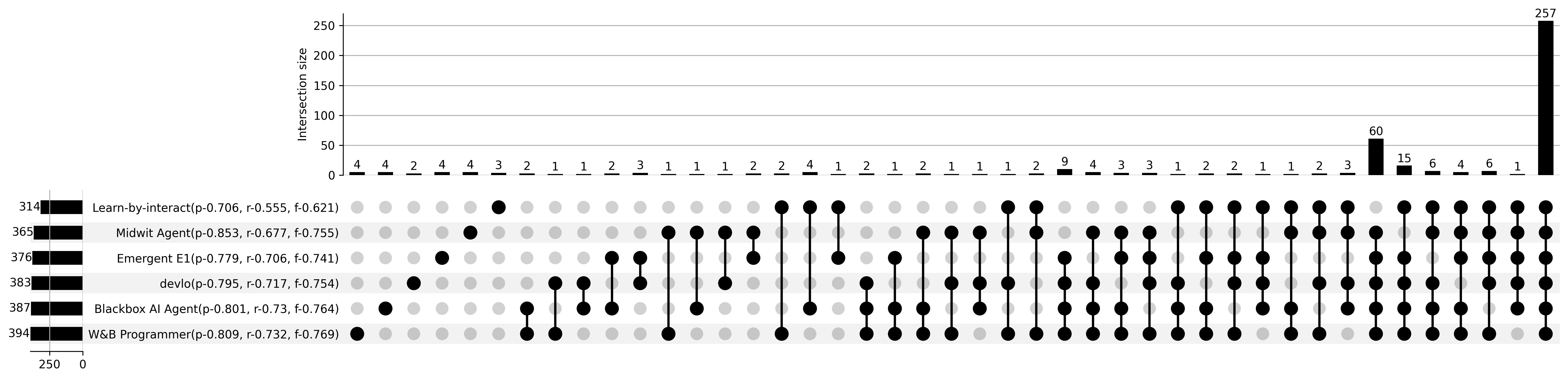}
    \caption{Statistics of samples where all buggy files are hit}
    \label{fig:file_fl_all_cover}
  \end{subfigure}
  
  \caption{Analysis of Performance of State-of-the-art Techniques on File-level FL Task}
  \label{fig:file_fl_analysis}
\end{figure*}

\begin{figure*}
  \centering
  \begin{subfigure}[b]{\textwidth}
    \includegraphics[width=\textwidth]{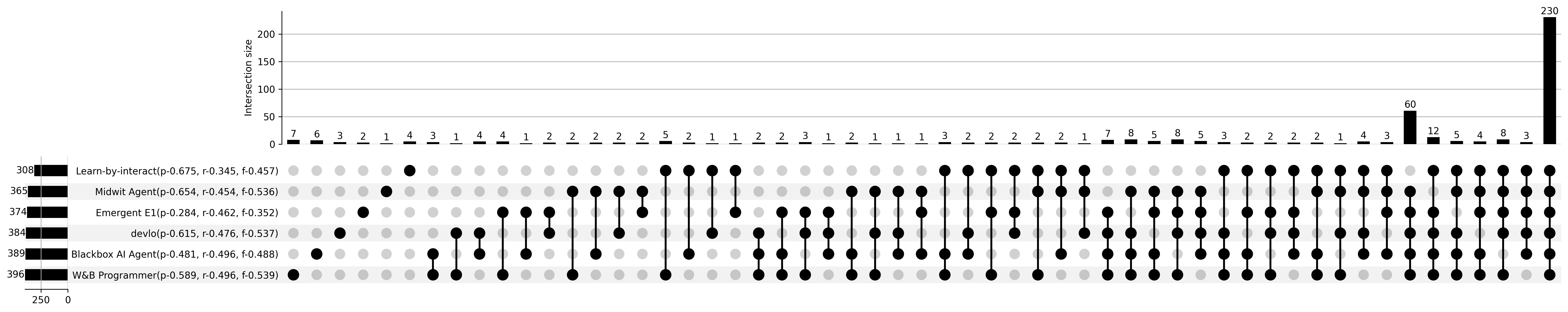}
    \caption{Statistics of samples where at least one buggy code symbol is hit}
    \label{fig:symbol_fl_intersection}
  \end{subfigure}
  
  \vspace{1em}  
  
  \begin{subfigure}[b]{\textwidth}
    \includegraphics[width=\textwidth]{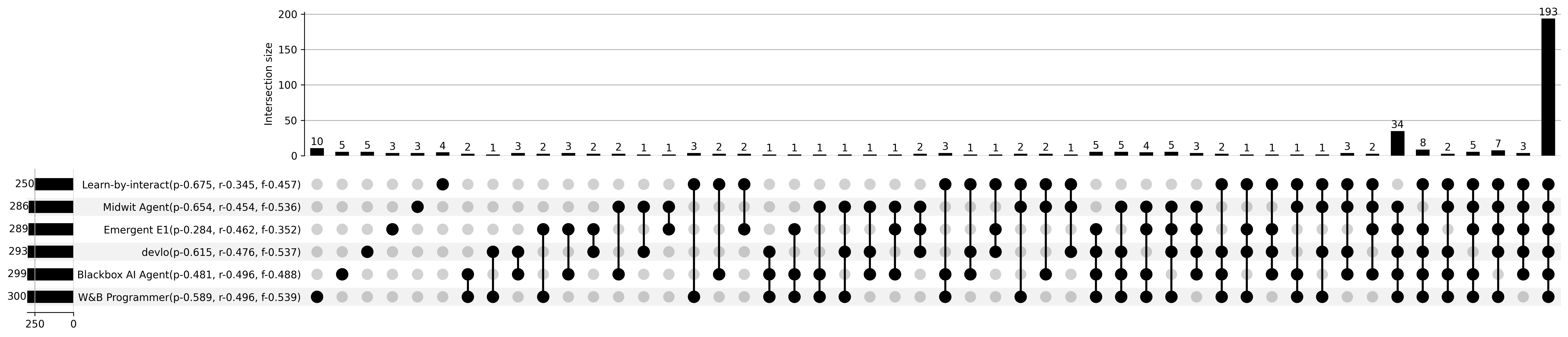}
    \caption{Statistics of samples where all buggy code symbols are hit}
    \label{fig:symbol_fl_all_cover}
  \end{subfigure}
  
  \caption{Analysis of Performance of State-of-the-art Techniques on Code Symbol-level FL Task}
  \label{fig:symbol_fl_analysis}
\end{figure*}

\newadd{To validate the generalizability of these observations, we further examined ten additional cases. We have to limit the scope of manual analysis due to two primary constraints. First, beyond certain score thresholds, issues no longer met the criteria for being considered extreme, reducing their diagnostic value. Second, each case required in-depth comprehension of the underlying codebase and task context, and the corresponding agent trajectories could span up to 200,000 tokens. This made detailed inspection resource-intensive. To ensure reliability, two authors independently reviewed each case and reconciled their assessments through discussion.}

\subsection{RQ2: Effectiveness of FL}

Fault localization (FL) is a critical step in resolving issues, and only when the faulty code element is accurately identified can the model generate a semantically correct patch. In this RQ, we examine the effectiveness of six methods in FL. As run trajectories of most repair techniques lack intermediate FL results or provide insufficient localization granularity (e.g., only offering file-level FL information, without finer-grained code symbol-level or line-level details), we extract the localization information in different levels from the final submitted patch to fairly assess each tool's localization capability. Specifically, we downloaded the final patch set submitted by each tool on the official SWE-bench Verified website, then parsed the formatted patch files. From the perspective of the files pre-modification (pre-patch code files), we recorded the files and code symbols that were modified. We did not conduct line-level experiments because we believe that individual lines of code do not adequately represent a complete functional module, and therefore, have limited reference value. We applied the same processing to the golden patches.

After obtaining the above processed data, \newadd{we created Figure~\ref{fig:file_fl_analysis} and Figure~\ref{fig:symbol_fl_analysis} to illustrate the hit rate of FL at the file and code symbol levels, respectively. The meaning of the different modules in the figures is similar to Figure~\ref{fig:repair_analysis}, except that Figure~\ref{fig:symbol_fl_analysis} is oriented towards the FL task.} 

Furthermore, we analyzed the localization performance of the six tools based on their definition of a hit. As shown in Figure~\ref{fig:file_fl_analysis}, the W\&B Programmer outperformed others in the file-level localization task, successfully localizing at least one buggy file in 448 cases (Figure~\ref{fig:file_fl_intersection}), and all buggy files in 394 cases (Figure~\ref{fig:file_fl_all_cover}). In the "Cover-at-least-one-file" setup, W\&B Programmer is followed by Devlo, Blackbox AI Agent, Emergent E1, Midwit Agent, and finally Learn-by-interact with 342 cases localized. This indicates a 31.0\% improvement for W\&B Programmer over Learn-by-interact. In the "Cover-all-files" scenario, only Devlo and Blackbox AI Agent swapped their positions, while the rest remained in the same order. The top-ranked tool, W\&B Programmer, showed a 25.5\% improvement over the lowest-ranked Learn-by-interact.

Additionally, since each instance may have multiple buggy files and each system may predict a list of files rather than a single file, we compiled the prediction accuracy, recall, and F1-score for each system at the code file level. This data is provided after the name of each system in Figure~\ref{fig:file_fl_analysis}. Based on the F1-score, the systems, sorted from lowest to highest, are Learn-by-interact, Emergent E1, Devlo, Midwit Agent, Blackbox AI Agent, and W\&B Programmer. Excluding Learn-by-interact, the relative order of the remaining systems aligns with their ranking in the issue resolution task, indicating a strong positive correlation between accurate file-level localization and overall bug-fixing performance.

Next, we assess FL performance at the code symbol level. In the "Cover-at-least-one-symbol" scenario, the top bug-fixing system is W\&B Programmer, successfully localizing 396 cases. The lowest bug-fixing system is Learn-by-interact, with 308 cases localized, resulting in a 28.6\% improvement for W\&B Programmer. In the "Cover-all-symbols" scenario, W\&B Programmer localized 50 more cases than Learn-by-interact, showing a 20\% improvement.
Compared to file-level results, code symbol-level localization shows more variance from the issue resolution task, and the abilities of various indicators differ. For example, W\&B Programmer ranks highest in F1-score for code symbol prediction, making it an all-around performer, while Midwit Agent excels in prediction precision. Although Blackbox AI Agent does not rank in the top three for precision and recall, it remains the second-best in localization from an instance perspective.

We believe these differences are primarily due to the varying localization strategies at the code symbol granularity employed by different models. Some repair systems might focus on recalling more code symbols to provide more flexibility for subsequent patch generation. Others might aim to precisely identify the first erroneous code symbols to avoid misguidance in subsequent repair processes, without overly considering the completeness of code symbol recall and associated modifications. Despite these differences, we still observe a positive correlation between code symbol-level FL results and bug-fixing results. Additionally, compared to the file-level, there is still considerable room for improvement in symbol-level localization effectiveness in the future.

It is easy to observe that in localization experiments, Learn-by-interact consistently ranks last and performs significantly below the other five repair systems. However, in the bug-fixing task, it ranks fourth. To investigate this, we constructed three case sets: 301 cases that Learn-by-interact successfully repaired (Set-Repair), 342 cases where it successfully localized at least one erroneous file (Set-File-Localization), and 308 cases where it localized at least one erroneous code symbol (Set-Symbol-Localization).

The intersection of Set-Repair and Set-File-Localization is 281 cases, indicating that if Learn-by-interact can successfully localize at least one erroneous file, the probability of ultimately resolving the case is $281/342 = 82.2\%$. The intersection of Set-Repair and Set-Symbol-Localization is 262 cases, showing that if it can successfully localize at least one erroneous code symbol, the probability of resolving the case is $262/308 = 85.1\%$. 

In comparison, these probabilities for W\&B Programmer are 69.6\% and 72.7\%, respectively. Thus, while Learn-by-interact's success rate in predicting the correct files and code symbols is minimal, once a prediction is successful, the subsequent repair success rate is remarkably high. This is why Learn-by-interact can achieve relatively good repair outcomes despite its poor FL performance.

\begin{tcolorbox}
 \textbf{Insights:} (7) Existing methods have achieved relatively strong performance in file-level FL, but there remains room for improvement in code symbol-level localization tasks. (8) Both coarse-grained file-level FL results and fine-grained code symbol-level FL results are positively correlated with final bug-fixing performance.
\end{tcolorbox}

\subsection{RQ3: Effectiveness of Bug Reproduction}

Under most of top-leading bug-fixing systems, the core process involves searching for suspicious codes, generating bug reproduction scripts, successfully reproducing the issue, and modifying the functional code until the issue is resolved (i.e., passing the reproduction testings). The generation of bug reproduction scripts is of paramount importance for the model's understanding of the problem and for verifying the correctness of the generated patches. Therefore, in this RQ, we plan to specifically investigate the significance of this task. An initial idea was to extract the reproduction scripts generated during the repair process from the trajectories submitted by the top six repair methods. However, we encountered numerous difficulties during the attempts.

Reproduction scripts within a project have variable names and locations, which complicates the accurate extraction of filenames using rule-based methods. These scripts may be named differently and stored in various directories, adding complexity. Additionally, trajectories do not directly provide the final versions of reproduction files, as repair methods often modify the initially visible contents using code editing tools. These modifications are partially visible, yet the full text remains obscured. The proprietary nature and diverse designs of these tools make understanding and fitting their functions challenging. Moreover, the execution commands for these files also vary. Some methods generate new Python files with a \texttt{main} entry point, while others alter existing test files and execute them using frameworks such as \texttt{pytest}. Extracting these commands from trajectories is thus a difficult task. Finally, problem reproduction processes may involve altering the running environment, such as installing dependencies via \texttt{pip} or exporting environment variables, which, if not recorded, can impede the execution of reproduction scripts effectively.

Having abandoned the idea of extracting and analyzing reproduction scripts from existing repair systems, we considered implementing an Agent with a standard process to preserve the final reproduced script during its runtime, thereby advancing subsequent investigations. Specifically, we designed a repair framework with two Agents, named RepoFixer. \newadd{Compared with other SOTA bug-fixing systems, for example, Agentless employs a single-step, phase-separated inference process, while SWE-agent uses a multi-step ReAct-based framework. Our implementation aligns more closely with the latter, but simplifies the agent architecture to retain only core components—avoiding heavy optimizations—to better generalize insights for bug reproduction tasks.} In RepoFixer, the first Agent, called Searcher, takes as input the issue description from each sample in the SWE-bench Verified dataset and the current commit state of the code repository. We instruct the Searcher via system prompt commands to perform a code symbol granularity FL task and, through multi-turn conversations, continuously search and read the context related to the issue description, ultimately returning a list of suspicious code symbols.

Our second Agent, called Fixer, takes as input the issue description for each sample and the list of suspicious code symbols returned by the Searcher, aiming to generate a patch that resolves the issue. In the system prompt, we suggest the Fixer's workflow: first, generate a problem reproduction script, then run it and self-assess whether the issue is successfully reproduced based on the run results. If the issue is not successfully reproduced, continue modifying the reproduction script until it is deemed that the issue described in the issue has been successfully reproduced. Then, generate the patch and use the reproduction script to verify whether the modified code successfully resolves the issue (i.e., the previously reproducible issue no longer occurs). If successful, save the patch and end the process; otherwise, continue to repair the patch and run the reproduction script until the issue is considered resolved or the predefined number of dialogue turns is reached. Regardless of whether a patch is successfully generated, we will save the latest edited reproduction script and the corresponding execution commands.

During this process, we utilized the powerful Claude 3.5 Sonnet model (October 2024 version)\footnote{\url{https://www.anthropic.com/news/3-5-models-and-computer-use}} as the base model to implement the Agent process. We used Claude's official toolkits, namely \texttt{bash} and \texttt{str\_replace\_editor}, as the Agent runtime toolkits and utilized the standard Function Call interface for model requests. Post-experimental statistics show that our RepoFixer successfully resolved a total of 255 issues, accounting for 51\% of all samples. While we did not adopt very heavy sampling or regression testing strategies, and complex mechanisms such as self-reflection, it appears that there remains some gap in performance compared to the top-ranked methods. However, the current focus of our research question is on investigating the quality of reproduction scripts generated by Agent systems. To explore this, we designed the following experiment.

We placed the collected reproduction scripts back into the Docker environment for each sample and then ran the script once in the repository before applying the golden patch, recording the output as \texttt{before\_fix\_status}. Subsequently, in a new Docker instance, we applied the golden patch to the code repository and then ran the same reproduction script again, recording the output as \texttt{after\_fix\_status}. The golden patch refers to the correct repair patches provided in the SWE-bench Verified dataset. A natural consideration is that if the output of the same reproduction script changes before and after applying the golden patch, it can be assumed that the reproduction script is related to the issue that the golden patch addresses. Thus, reproduction scripts producing such status differences should be considered of higher quality, whereas those without status differences seem likely to have a weaker association with the issue.

Based on this, we collected experimental results. Out of 500 instances generating 500 reproduction scripts, 384 scripts produced status differences before and after the golden patch application to the repository, representing 76.8\%. We term reproduction scripts exhibiting this property as "related reproduction scripts." We hypothesize that the higher the quality of the reproduction information in the issue description, the more likely the model is to generate related reproduction scripts. For this, we conducted detailed statistics. Among 194 samples containing "Contains REs," 156 produced related reproduction scripts, accounting for 80.4\%; among 169 samples containing "Contains Partial REs," 135 produced related reproduction scripts, accounting for 79.9\%; among 87 samples where the information is in natural language ("Info in NL"), 61 produced related reproduction scripts, accounting for 70.1\%; and among 50 samples marked "Not Enough Info," 32 produced related reproduction scripts, accounting for 64\%. It can be observed that as the quality of reproduction descriptions in the issue decreases, the proportion of generating related reproduction scripts also decreases. Thus, for users, accurately and comprehensively recording the process of triggering an issue in code form is of significant value for automating the resolution of the issue.

\begin{tcolorbox}
 \textbf{Insights:} (9) The most effective bug-fixing systems currently employ a process based on bug reproduction, patch generation, and patch correctness verification. Among these, the bug reproduction process holds significant value. (10) Drafting clear and comprehensive bug reproduction cases within the issue can greatly reduce the difficulty of reproducing the bug, consequently enhancing the efficiency of end-to-end bug-fixing tasks.
\end{tcolorbox}

\section{Discussion}
\label{sec:discussion}
\subsection{Large Language Model}
From the LLM perspective, it is necessary to further enhance the model's reasoning ability so that it can accurately identify information related to the bug within the issue, thereby reducing the interference of noise. Additionally, for multiple potential repair locations, the model should utilize its reasoning capability to select the location most relevant to the issue.

\subsection{Agentic Flow}
Agents should especially focus on the quality of the issue and pay attention to multiple suspicious locations in the stack trace. The Agentic flow design should include mechanisms to check the completeness of patches and consider the global impact of the fixes. During the use of the model, mechanisms should be established to either avoid the randomness of the model's output or make full use of the diversity in the model's output. The effectiveness of fault localization is important for bug-fixing tasks, finer-grained code symbol level fault localization task still has significant room for improvement. During the reproduction process, it is necessary to increase the proportion of the dynamic repair process, centered on bug reproduction, patch generation, and patch validation. The quality of bug reproduction scripts should be improved by introducing more relevant artifacts and increasing model reflection.

\subsection{General Applicability}

\newadd{While our study uses the SWE-bench Verified benchmark with only Python repositories, we believe the analytical insights generalize beyond this setting.
The benchmark is widely used by organizations such as OpenAI and Anthropic, and the analyzed systems rely on language-agnostic tools, with no signs of Python-specific tuning. Nevertheless, we recognize the importance of multilingual benchmarks and are developing a broader dataset for future evaluations.}


\subsection{Qualitative Evaluations from Developers}

\newadd{While our evaluated patch correctness via test suite validation, 
we acknowledge it may not fully capture patch quality from a developer’s perspective. To supplement this, we manually analyzed 30 patches per system, evaluating code style, readability, and adherence to development norms. We found that agent-based workflows enable context-aware understanding, produce syntactically correct code, reuse existing structures effectively, and often yield clean, human-readable patches with meaningful comments}



\newadd{These results indicate that modern LLM-based agents can generate patches that are both correct and stylistically aligned with professional standards. A more comprehensive evaluation with developer feedback remains future work.}

\section{Threats to Validity}
\label{sec:threats}
\noindent\textbf{Fail-to-Pass Tests:}
SWE-bench uses Fail-to-Pass (F2P) tests to verify the correctness of generated patches. However, F2P tests may not be comprehensive, allowing a patch to pass F2P and be deemed correct without fully addressing the user's issue. This is a common problem in the field of APR as well as in LLM evaluation based on unit tests~\cite{liu2024your}. In this context, we assume that a patch is correct as long as it passes the F2P test cases. We also call for contributions from the academic community to improve the test cases in the SWE-bench evaluation dataset to make the evaluation results more reliable.

\noindent\textbf{Uncertainty of LLM:}
The output of LLMs is stochastic, leading to a probabilistic nature for whether an instance is solved. 
In this work, we directly analyzed the patches and trajectories submitted by various systems, assuming that the results submitted to SWE-bench represent the best performance of the Agent. 
Furthermore, conducting multiple experiments for each system to eliminate stochasticity is impractical in terms of both cost and accessibility. \newadd{As these systems are typically closed-source, we cannot standardize their model and decoding configurations or conduct a controlled ablation using identical LLM backbones. Thus, differences in model capacity inherently contribute to variations in performance, a factor we acknowledge explicitly as a limitation of our comparative analysis.}

\noindent\textbf{LLM-based Issue Quality Assessment:} To ensure that the LLM's evaluation results are as close as possible to those of software experts, we took the following measures: (1) We selected the DeepSeek-R1 model, which has top-tier inference performance, as the evaluator; (2) Initially, we conducted model annotations and manual annotations on a subset of the SWE-bench Verified dataset, continuously refining the scoring prompts based on discrepancies until they aligned with human expectations; (3) We used DeepSeek-R1 to perform five annotations for each metric of each instance and selected the option with the highest frequency as the final choice.

\noindent\textbf{Reliability of Reproduction Scripts:} \newadd{
Another potential threat to validity lies in the possibility of false positives—scripts marked as “related” that do not actually trigger the bug. While our approach leverages output divergence as an indication of successful bug reproduction, inspired by mutation-based fault localization~\cite{li2017transforming}, this method may still misclassify some non-triggering scripts as successful reproductions. Incorporating additional checks, such as verifying log patterns or exception types, could theoretically reduce such misclassification. However, given the practical complexity of implementing and reliably automating these checks within limited agent interactions, we consciously accepted this limitation to maintain a balance between practicality and effectiveness. Future work could explore incorporating such checks to more precisely measure reproduction accuracy and further mitigate this validity threat.
}

\section{Related Work}
\label{sec:related}

\subsection{Large Language Models}

Large language models (LLMs) are highly advanced pre-trained language models.
These models undergo initial unsupervised training on vast amounts of corpus, followed by fine-tuning for specific tasks to enhance performance.

Language models are classified into three categories based on their architecture: encoder-only models~\cite{feng2020codebert}, decoder-only models~\cite{nijkamp2022codegen}, and encoder-decoder models~\cite{tian2022learning}.
Most existing LLMs for code utilize the transformer architecture's encoders, known for their exceptional learning capabilities and scalability.
Regardless of their architecture, most models can be fine-tuned with task-specific data to enhance performance~\cite{lin2023cct5}.

Large language models (LLMs) have become a promising choice for various software engineering tasks due to their impressive performance in both code generation and understanding~\cite{yang2023enhancing}, such as program synthesis~\cite{liu2024your, zhu2024sketch, wang2023practitioners, wang2023two, wang2023natural}, code translation~\cite{yu2024Mal, yang2024exploring}, program repair~\cite{lin2024one, jiang2023impact, xia2023automated}, fault detection and localization~\cite{du2024generalization, qin2024agentfl}, incident analysis~\cite{chen2024automatic, ahmed2023recommending}, code summarization~\cite{geng2024large} and testing~\cite{sun2023smt}.
For example, Codex~\cite{chen2021evaluating}, StarCoder~\cite{lozhkov2024starcoder}, and DeepSeek-Coder~\cite{zhu2024deepseek} are notable code-specific LLMs developed through extensive training on large datasets of open-source code snippets.
Additionally, instruction-following code-specific LLMs such as DeepSeek-Coder-Instruct~\cite{zhu2024deepseek} and Magicoder~\cite{wei2023magicoder} have been created using instruction-tuning methods to enhance their utility in coding tasks.

\subsection{Fault Localization and Program Repair}

Fault localization (FL)~\cite{wong2016survey} techniques aim to discover and analyze the location and causes of faults, which can be categorized into dynamic and static approaches.
Dynamic techniques, such as spectrum-based fault localization (SBFL)~\cite{abreu2007accuracy, abreu2009practical} and mutation-based fault localization (MBFL)~\cite{papadakis2015metallaxis}, analyze the dynamic execution information to determine fault locations, though they are resource-intensive.
Static FL techniques~\cite{mao2014slice} determine fault locations through semantic or syntactic analysis at the bug report or source code level, offering fast detection with low resource consumption.
Advanced FL techniques, such as multiple fault localization (MFL) and combined dynamic and static methods, have emerged to find more errors~\cite{xiao2021albfl, kim2019precise, neelofar2017improving}.


Automated program repair (APR)~\cite{le2019automated} has attracted significant attention over the past decade.
APR techniques aim to generate patches for buggy programs to pass given test suites.
These techniques can be categorized into search-based~\cite{li2022improving, mehne2018accelerating}, semantics-based~\cite{le2017jfix, nguyen2013semfix, le2016empirical}, and pattern/learning-based approaches~\cite{li2020dlfix, li2022dear, zhang2023survey}.
Search-based APR techniques like GenProg~\cite{le2011genprog} use predefined code mutation operators to generate patches, while semantics-based APR techniques generate patches by solving repair constraints based on test suite specifications.
Learning-based APR techniques, such as those utilizing deep learning models, train on large code repositories to predict correct patches.
Recent work uses LLMs for APR, often focusing on constructing APR-specific prompts to guide LLMs in generating patches for buggy program statements~\cite{xia2023automated}.

\subsection{Agents for Software Development}

The emergence and popularity of agent-based frameworks have led to the development of agent-based approaches for solving software engineering tasks.
Devin and its open-source counterpart OpenDevin~\cite{wang2024opendevin} are among the first end-to-end LLM agent-based frameworks.
These frameworks use agents for planning based on user requirements and enable agents to iteratively perform tasks using tools like file editors, terminals, and web search engines.
SWE-agent~\cite{yang2024swe}, for example, designs a custom agent-computer interface (ACI) allowing LLM agents to interact with the repository environment through actions such as reading, editing files, and running bash commands.
AutoCodeRover~\cite{zhang2024autocoderover} provides LLM agents with specific APIs to effectively identify locations needing modification to resolve issues. Numerous other agent-based approaches have been developed, both in open-source and commercial products.



\section{Conclusion}
\label{sec:conclusion}
In this paper, we analyzed the top 6 bug-fixing systems on SWE-bench Verified. We conducted detailed analyses of the performance of LLM-based Agents in automatic bug fixing for code repositories, and the performance of different systems in different levels of fault localization. Moreover, we developed a mainstream workflow bug-fixing agent to evaluate the overall performance on bug reproduction tasks. The analysis results indicate that to further enhance the capabilities of LLM-based Agents in bug fixing, future efforts should focus on improving the reasoning ability of LLMs, and the Agentic flow design, considering the quality of issues, stack traces, and the correctness of reproductions.

\section*{Data availability}
The code and related experimental data in this paper are accessible on this page: \textit{\url{https://github.com/ResearchOpenRepos/bug_fixing_agent_empirical_study}}

\bibliographystyle{ACM-Reference-Format}
\bibliography{references}

\end{document}